\documentclass[12pt]{article}

\newread\testifexists
\def\GetIfExists #1 {\immediate\openin\testifexists=#1
    \ifeof\testifexists\immediate\closein\testifexists\else
    \immediate\closein\testifexists\input #1\fi}

\usepackage{gthstyle}
\immediate\openin\testifexists=e
    \ifeof\testifexists\immediate\closein\testifexists\else
    \immediate\closein\testifexists\input e\fipsf

\def\Bbb#1{\setbox0=\hbox{$\tt #1$}  \copy0\kern-\wd0\kern .1em\copy0}

\def\bbf#1{\setbox0=\hbox{$#1$} \kern-.025em\copy0\kern-\wd0
        \kern.05em\copy0\kern-\wd0 \kern-.025em\raise.0433em\box0}

\immediate\openin\testifexists=a
    \ifeof\testifexists\immediate\closein\testifexists\else
    \immediate\closein\testifexists\input a\fimssym.def          

      \def\b{\beta}         
        \def\e{\varepsilon}
          \def\l{\lambda}     \def\L{\Lambda}
\def\m{\mu}                     \def\vv{\varphi}
\def\n{\nu}         \def\j{\psi}    
\def\r{\varrho}       \def\SS{\Sigma}
        \def\th{\theta}

\def\HH{{\cal H}} \def\LL{{\cal L}} \def\OO{{\cal O}}
\def\pa{\partial} \def\ra{\rightarrow}

\def\dd{{\rm d}}

\def\fract#1#2{{\textstyle{#1\over#2}}}
\def\ffract#1#2{\raise .3 em\hbox{$\scriptstyle#1$}\kern-.25em/
                \kern-.2em\lower .2 em \hbox{$\scriptstyle#2$}}

\def\half{\fract12} \def\quart{\fract14} 

\def\part#1#2{{\partial#1\over\partial#2}}

\def\iss{\ =\ }

\newcommand{\tl}[1]{\tilde{#1}}

\newcommand{\be}{\begin{eqnarray}}
\newcommand{\ee}{\end{eqnarray}}
\newcommand{\eqn}[1]{(\ref{#1})}
\newcommand{\nn}{\nonumber\\}

\newcommand{\bi}[1]{\begin{itemize}\item[#1]}

\newcommand{\ei}{\end{itemize}}

\newcommand{\fn}{\footnote}

\newcommand{\newsec}[1]{\section{#1}\setcounter{equation}{0}}

 \newcommand{\eel}[1]{\label{#1}\end{eqnarray}}\newcommand{\crl}[1]{\label{#1}\\ }

\begin{document}

\begin{titlepage}

\title{\normalsize \hfill ITP-UU-04/18  \\ \hfill SPIN-04/11
\\ \hfill {\tt hep-th/0408183}\\ \vskip 20mm \Large\bf
CONFINEMENT AT LARGE \(N_c\)
\thanks{Presented at \emph{Large $N_c$ QCD}, Trento, July 5-9,
2004}}

\author{Gerard 't~Hooft}
\date{\normalsize Institute for Theoretical Physics \\
Utrecht University, Leuvenlaan 4\\ 3584 CE Utrecht, the
Netherlands ??\medskip \\ and
\medskip \\ Spinoza Institute \\ Postbox 80.195 \\ 3508 TD
Utrecht, the Netherlands \smallskip \\ e-mail: \tt
g.thooft@phys.uu.nl \\ internet: \tt
http://www.phys.uu.nl/\~{}thooft/}

\maketitle

\begin{quotation} \noindent {\large\bf Abstract } \medskip \\
A discussion is given of the confinement mechanism in terms of the
Abelian projection scheme, for a general number \(N_c\) of colors.
There is a difficulty in the \(N_c\ra\infty\) limit that requires
a careful treatment, as the charges of the condensing magnetic
monopoles tend to infinity. We suggest that Bose condensation of
electric or magnetic charges is indicative for the kind of
confinement that takes place, but the actual mechanism of
confinement depends on other features as well.
\end{quotation}

\vfill \flushleft{\today}

\end{titlepage}

\eject

\newsec{Introduction: The \(\b\) function.}
In the absence of mass terms, field theories used in particle
physics often appear to be scale-independent. As is well-known,
however, quantization and renormalization of these theories
require a scale-dependent cut-off, and the scale dependence in
general does not go away in the limit where one sends the cut-off
to infinity.\cite{CS} If \(\m\) is the average value of the
momenta\fn{Often, the \(\b\) function is defined to refer to
\(\m^2\dd/\dd\m^2\) of some coupling strength, which leads to a
factor 2 in Eq.~\eqn{betaQED}.} in an amplitude that is computed
perturbatively, and if the subtractions are carried out such that
the higher-order corrections for this amplitude are kept as small
as possible (in order to obtain a reasonably convergent
perturbation expansion), then one finds the coupling parameters
\(g\) to be \(\m\)-dependent. In case of QED, one finds the
electric charge parameter \(e\) to obey\cite{G-ML}
\be{\m\,\dd\over\dd\m}e^2(\m)=\b(e^2)\ ,\qquad \b(e^2)={e^4\over
6\pi^2}N_f\ +\ \OO(e^6)\ . \eel{betaQED} The dominant contribution
to \(\b(e^2)\) comes from the one-loop diagram in the photon
propagator, see Fig.~\ref{QEDYMbeta.fig}. It is proportional to
the number \(N_f\) of charged fermion species. If charged scalar
fields are present they also contribute to \(\b\), with the same
sign.

\begin{figure}[h] \setcounter{figure}{0}
\begin{quotation}
 \epsfxsize=130 mm\epsfbox{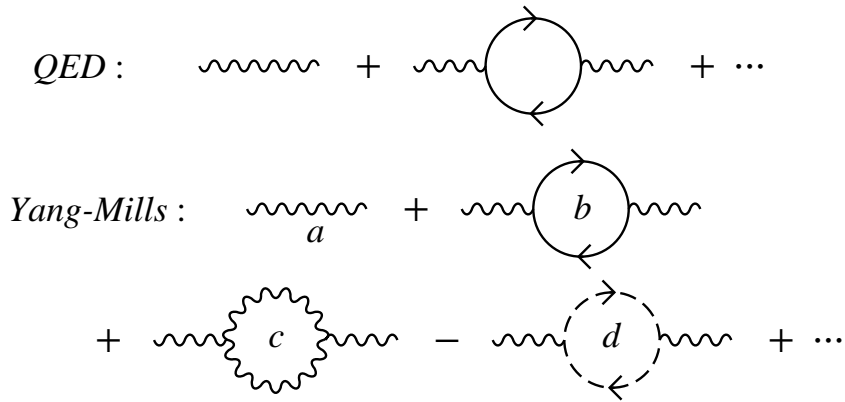}
  \caption{\small{Dominant contribution to the \(\b\) function for QED and for
  Yang-Mills.}}
  \label{QEDYMbeta.fig}\end{quotation}
\end{figure}

Before 1970, it was generally believed that all \(\b\) functions
in quantum field theories had to be positive. In the Yang-Mills
case, however, different results were
found\cite{GrossWil}\cite{GtHrengr}. The contribution of fermions
to \(\b\) is as in the QED case (see Fig.~\ref{QEDYMbeta.fig},
diagram ($b$). The contribution of the gauge bosons themselves,
however, is of the opposite sign (Fig.~\ref{QEDYMbeta.fig},
diagram ($c$)). When doing the calculation, in a convenient choice
of gauge, one finds a primary term that is like the contribution
of scalar particles, with relative strength \(+\fract16\), but in
addition a much larger contribution, of relative strength \(-1\),
from those terms that generate the \emph{magnetic moments} of the
gauge bosons. These are large, since they have spin one and
giromagnetic ratio 2. Finally, there is a small negative
contribution from the ghosts, diagram ($d$), of strength
\(-\fract1{12}\). The net result is \be \b(g^2)={g^4\over
12\pi^2}(N_f-11)+\OO(g^6)\ ,\quad\mathrm{for }\ SU(2)\
;\crl{SU2beta} \b(g^2)={g^4\over 6\pi^2}(2N_f-11N_c)+\OO(g^6
N_c^2)\ ,\quad\mathrm{for }\ SU(N_c)\ ,\quad N_c>2\ .
\eel{SU3beta} The \(SU(2)\) case deviates only because, there, the
usual definition of the color coupling \(g\) is chosen with a
factor \(\half\).

Ignoring the higher order terms, the solution of these equations
for the running coupling parameter is \be
g^2(\m)=-{1\over\b_2\,\ln(\m/\L)}\ , \eel{running} where \(\b_2\)
is the coefficient in front of the \(g^4\) term in the expansion
for \(\b\), and \(\L\) the fundamental scale parameter of the
theory. If \(\b_2\) is negative, \(\L\) has to be taken small, and
perturbation expansion only makes sense at \(\m\gg\L\). Then \(g\)
tends to zero at large \(\m\), but it explodes as
\(\m\downarrow\L\).

Does this behavior of the running coupling parameter for
Yang-Mills theories such as QCD imply a permanently confining
force between quarks? Today, this is indeed believed to be a quite
natural consequence, but in the 1970's, the problem of
completeness was brought up. What does the spectrum of physical
states look like? if we exclude free quarks and free gluons, can
we then ever establish unitarity of the scattering matrix? The
only way to understand how unitarity can be restored, is to view
confinement as a new phase of matter. It is related to topological
features of the gauge theory.

\newsec{Magnetic and electric confinement} \subsection{Magnetic
confinement} The first sign of an absolutely confining force
emerging in a conventional quantum field theory, came from the
study of the Abelian Higgs theory\cite{NO}. Take the Lagrangian
\be\LL(A,\vv)=\quart F_{\m\n}F_{\m\n} - D_\m\vv^\dagger
D_\m\vv-V(\vv)\ , \eel{AbelHiggs} where \(\vv\) is a single,
complex scalar field, and \(V(\vv)\) a quartic potential invariant
under complex rotations of \(\vv\): \be
V(\vv)=\half\l(\vv^\dagger\vv-F^2)^2\ . \eel{Vphi} Here, \(F\) is
a fixed parameter. The physical vacuum is described by \(\vv\)
staying close to its equilibrium value: \(\vv=Fe^{i\th}\), where
\(\th\) may be arbitrary. \(\th\) is fundamentally unobservable
since it is completely gauge-dependent.

If, however, in a `sheet', that is, two-dimensional subspace of
space or space-time, \(\th\) rotates over a full \(360^\circ\),
then the \(\vv\) field develops a `frustration': \(\vv\) must be
differentiable, because of the derivative terms in \(\LL\), and
therefore there must be a zero somewhere in the sheet. Moving the
sheet along in space, we find that this zero forms a one
dimensional line in 3-space, i.e., a vortex. In the immediate
vicinity of this vortex, \(\vv\) deviates considerably from its
equilibrium value, so that the vortex will carry energy\fn{This
extra energy is shared with that of the kinetic term for the
\(\vv\) field in \eqn{AbelHiggs}, that enforces continuity of
\(\vv\).}. Away from the vortex, the equilibrium value \(F\) (or a
rotation thereof) is quickly resumed, and so, the vortex maintains
a finite transverse extension. It is a non-trivial, locally stable
field configuration.

Some elementary calculations show that this vortex carries
magnetic flux. Therefore, if we take a magnetic monopole and its
antiparticle, i.e., a north and a south pole, then they will be
connected by a vortex, causing an absolutely confining force
between them, since the energy is proportional to the vortex'
length.

This phenomenon  by itself is not new; it was known to describe
the Meissner effect in super-conducting materials. Now we see that
it leads to the existence of magnetic vortex lines in the vacuum
of the Higgs theory. The magnetic confinement model of this
section would only explain confinement of quarks if quarks carried
a magnetic monopole charge. It was once thought that quarks indeed
carry magnetic monopole charges.

\subsection{Electric confinement}
This, however, is not the case in QCD; quarks only carry a
color-electric monopole charge. Thus, what is needed to understand
confinement of quarks is the description of color-electric vortex
lines. These are related to the magnetic vortex lines by a
\emph{dual transformation}\cite{GtHMandel}: \(\vec E\ra\vec B\),
\(\vec B\ra-\vec E\). This leaves the homogeneous parts of the
Maxwell equations invariant, but replaces electric charges with
magnetic ones and \textit{vice-versa}. Since magnetic monopole
charges do tend to occur in non-Abelian gauge theories, one may
suspect the occurrence of \emph{magnetic super-conductivity}: the
magnetic monopoles condense.

On the other hand, we must keep in mind that stable magnetic
monopoles only seem to occur in theories where a compact gauge
group is spontaneously broken into a surviving \(U(1)\) subgroup.
How can we follow the activities of `monopoles' if the symmetry is
not spontaneously broken, as in QCD?

 \newsec{The Abelian projection for general
 \(N_c\)}\label{Abelproj.sec}
 Apart from the commutator terms in the Lagrangian, there is
another fundamental difference between Abelian and non-Abelian
gauge theories. In Abelian theories, it is impossible to fix the
gauge locally, at some space-time point \(x\), without referring
to the field configurations at other space-time points, far away
from \(x\), unless one uses charged scalar fields that must have
been added to the system. In a \emph{non}-Abelian gauge theory,
one can fix the \emph{non-Abelian part} of the gauge redundancy by
referring exclusively to the vector potential and at most its
first derivatives, at the point \(x\) alone. This means that,
without adding non-local elements to the Lagrangian, one can
rewrite a non-Abelian gauge theory as if it were an Abelian one.
The only price one pays is that the new, Abelian, Lagrangian
becomes a non-polynomial one. The new gauge group is the
\emph{Cartan sub-group} of the original non-Abelian gauge group.

We call this procedure the \emph{Abelian
projection}\cite{GtHAbel}. In what follows, we describe it for
\(SU(N)\) for general \(N\). The Cartan sub-group of \(SU(N)\) is
\be {(U(1))^N\over U(1)}\iss U(1)^{N-1}\ \subset\ SU(N)\ .
\eel{Cartan} Take any component of the (non-Abelian) field tensor,
say \({G_{12}}^i_{\,j}\). Here, \(i\) and \(j\) are gauge indices
running from 1 to \(N\). By selecting out the 12\,-\,direction in
Minkowski space, our gauge choice will violate Lorentz invariance.
It is not really necessary to break Lorentz invariance; one could
have chosen any Lorentz-invariant hermitean matrix constructed
from the \(G_{\m\n}\), but this would be technically more
complicated, and no harm is done with our simpler choice.

An Abelian projection is realized by choosing the gauge in which
\(G_{12}\) is diagonalized:\be
G_{12}(x)=\pmatrix{\l_1(x)&0&\cdots& 0\cr 0&\l_2(x)&\cdots&0\cr
\vdots&\vdots&\ddots&\vdots\cr 0&0&\cdots&\l_N(x)}\
,\quad\l_1\le\l_2\le\cdots\le\l_N\ . \eel{Aproj} Indeed, we made
use of the \(SU(N)\) subgroup of the pure permutations to order
the eigenvalues \(\l_i(x)\).

Note that, even if the Jacobian associated with the
transformation from the vector fields \(A_\m(x)\) to the fields
\(\l_i(x)\) may be non-trivial, there are no ghosts associated
with it. This is because the transformation is a local one: the
Faddeev-Popov field does not have a kinetic term. In this gauge,
all off-diagonal field components are physically significant ---
they are invariant under the remaining (Abelian) \(U(1)^{N-1}\) -
gauge transformations. Therefore, there are no massless, charged
vector bosons. The diagonal components of the photon fields do
survive as \(N-1\) different species of neutral, massless photons.

The fields which in the original Lagrangian came in the
fundamental representation, now split up into \(N\) different
fields. Their charges with respect to the \(N\) subgroups \(U(1)\)
can be labelled as \be \vec
Q=(0,\,\cdots,\,0,\,q,\,0,\,\cdots,\,0)\ . \eel{fundcharges} This
formula must be understood as describing the coupling to \(N\)
photons, which themselves are mixed in such a way that the
diagonal, `baryonic' \(U(1)\) photon is removed from the spectrum
of photons, so that \(N-1\) independent photon states survive. The
quark field component \(\j_i(x)\) is coupled, with charge \(q\),
to the \(i^\mathrm{th}\) photon.

The charged gluon fields do (partly) survive. Their charge table
is \be\vec
Q=(0,\,\cdots,0,\,q,\,0,\,\cdots,\,0,\,-q,\,0,,\,\cdots,\,0)\ .
\eel{gluoncharges} Note that they will not couple to the baryonic
\(U(1)\) - photon.

One would conclude that the emerging scheme is exactly as if we
had \(N-1\) ordinary Maxwell fields, coupled to particles with
various combinations of (Abelian) charges. There is, however, one
novelty: the Abelian projection is singular whenever two
eigenvalues \(\l_i(x)\) at a given point \(x\) coincide. Since the
\(\l_i\) were ordered, only two consecutive ones can coincide. In
the immediate neighborhood of such a point, the original field
\(G_{12}(x)\) takes the form \be
G_{12}(x)\simeq\pmatrix{*&|&0&|&*\cr\hline\cr 0&\bigg|&
\matrix{\l_0+a_3& a_1-ia_2\cr a_1+ia_2&\l_0-a_3}&\bigg|&0\cr
\hline\cr *&|&0&|&*}\ . \eel{hedgehog} the two consecutive
\(\l\)'s only coincide if \(a_1,\ a_2,\) and \(a_3\) all vanish.
these three conditions define isolated points in three-space.
Indeed, these points have the same characteristics as a magnetic
monopole in a Higgs theory with Higgs in the adjoint
representation of one of the subgroups \(SU(2)\) of \(SU(N)\).
Thus, at such points we find magnetic monopoles. Aparently, this
is the way the non-Abelian \(SU(N)\) theory differs fundamentally
from just any Abelian \(U(1)^{N-1}\) theory: besides the electric
charges of the form \eqn{fundcharges} and \eqn{gluoncharges}, we
have magnetic monopoles. With respect to the subgroup \(SU(2)\)
mentioned above, the monopole charge is in the Abelian subgroup
\(U(1)\) of \(SU(2)\). This means that the magnetic charge table
for the monopole is \be \vec g_m=(0,\,\cdots,\,0,\,g_m,\,-g_m,\,0,
\cdots,\,0)\ ;\qquad g_m=2\pi/q \eel{magncharges} (the subscript
\(m\) referring to `magnetic'). Note that the quarks obey the
minimal Dirac condition \be\sum_{i=1}^N q_i g_{mi}=2\pi n\ ,
\eel{Diraccond} with \(n=1\) or 0, while the charged gluon whose
charges are in the same \(SU(2)\) subgroup has \(n=\pm2\). We see
from the table \eqn{magncharges}, that there are \(N-1\) monopole
types.

``Confinement" now occurs in the following way. The
\(k^\mathrm{th}\) monopole field condenses to cause confinement
with respect to the Abelian subgroup \(U(1)_k\otimes
[U(1)_{k+1}]^{-1}\) of \(SU(N)\). This means that a vortex emerges
that confines charges in \(U(1)_k\) \emph{or} anti-charges in
\(U(1)_{k+1}\), by binding them to anti-charges in \(U(1)_k\)
\emph{or} charges in \(U(1)_{k+1}\). thus, the \(k^\mathrm{th}\)
monopole allows `hadrons' of the type \(q_i\,\overline q_i\) but
also `hadrons of the type \(q_k\,q_{k+1}\). in other words,
either all \(U(1)\) charges are neutralized, or the
\(k^\mathrm{th}\) charge must be equal to the
\((k+1)^\mathrm{th}\) charge.

The latter might seem to be an odd type of hadron, but we have to
realize that the \(k^\mathrm{th}\) monopole does not care about
the charges in other channels, and consequently, the
\emph{collective} action of all \(N-1\) monopole fields allow
only \(q_i\,\overline q_i\) objects, or objects where \emph{all}
charges are equal: \(\vec Q=(q,\,\cdots,\,q)\) to survive as
unconfined particles. The latter are the baryons.

We see that, unlike what one would expect, \(q_1\,\overline q_1\)
is not confined, whereas what one would expect is that only
\(\sum_{i=1}^N q_i\,\overline q_i\) would survive. This, however,
is a special feature of our gauge choice: the individual fields
\(q_i\,\overline q_i\) are indeed gauge-invariant here; we claim
that the Abelian projection does yield an accurate description of
the spectrum of mesonic states, even if it does not look very
realistic; the different states \(q_1\overline q_1,\ q_2\overline
q_2\), etc., are probably strongly mixed.

\newsec{Confinement and Bose condensation of charges} In the
\(N\ra\infty\) limit, one wishes to rescale the coupling
strengths: \be q={\tilde q\over\sqrt N}\ ;\qquad g_m=\sqrt
N\,\tilde g_m\ ;\qquad \tilde q\,\tilde g_m=2\pi\ .
\eel{tHcoupling} This is certainly also what is suggested by the
\(\b\) function Eq.~\eqn{SU3beta}. Consequently, the monopole
charge \(g_m\) itself tends to infinity. This makes the arguments
discussed above suspect; the methods of Quantum Field Theories
cannot be used to describe the Bose condensation of very strongly
interacting fields. Nevertheless, the \(\b\) function of
Eq.~\eqn{SU3beta} makes one believe that confinement continues to
take place as \(N\ra\infty\). Indeed, the planar diagrams in this
limit remind us of string diagrams, which have confinement built
in.

\subsection{\emph{Intermezzo}. Confinement as a universal laws in
the non-Abelian sector}\label{Intermezzo.sub} In view of the
above, one may formulate a conjecture that should hold for
\emph{all} non-Abelian gauge theories: \begin{quotation} For all
gauge groups \emph{except} \(U(1)\), all physical states are color
singlets.\end{quotation} Thus, we claim that magnetic monopoles
are not needed to achieve confinement, though they do provide for
a very useful signal: their vacuum expectation value. To
illustrate the point, let us give an unusual, but totally correct
description of the physical particles in the \(SU(2)\) sector of
the Standard Model.

The fermion doublet, \(\j_L\), the quark fields \(q_L\), the gauge
vector potential \(A_\m\), and the Higgs field \(\vv_H\) are
usually described as \be&&\j_L=\pmatrix{\n_{e,L}\cr e_L}\ ;\qquad
q_L=\pmatrix{u_L\cr d_L}\ ;\qquad A_\m=\pmatrix{W_\m^+\cr
Z_\m^0\cr W_\m^-}\ ;\nn &&\vv_H=\pmatrix{F\cr 0}+\tilde\vv\ .
\eel{SU2fields} However, we can describe all of the physical
fields as singlets. The fields \(\j_L\), \(q_L\), \(\vv\) and
\(A_\m\) are handled as \(SU(2)\)-quarks and gluons. Apart from
renormalization factors and tiny higher order corrections, the
\(SU(2)\)-mesons are \be \n_{e,L}=(\vv_H^*\cdot\j_L)\ ,\qquad
u_L=(\vv_H^*\cdot q_L)\ ,\qquad Z^0_\m=(\vv_H^*\cdot D_\m\vv_H)\ ,
\eel{SU2mesons} (the latter being ``\(P\) bound states"). The
\(SU(2)\)-baryons are \be e_L=\e_{ij}\,\vv_H^i\,\j^j_L\ ,\qquad
d_L=\e_{ij}\,\vv_H^i\,q_L^j\ ;\qquad
W^-_\m=\e_{ij}\,\vv_H^i\,D_\m\vv_H^j\ , \eel{SU2baryons} and
anti-baryons are constructed similarly. The only difference with
QCD is that, here, one can use conventional perturbation expansion
to calculate the properties of these particles in the usual way,
using the vacuum form \eqn{SU2fields} for the Higgs field.

Thus, we see that, using a somewhat unconventional language, the
Standard Model can be dealt with in such a way that \emph{both}
the \(SU(3)\) \emph{and} the \(SU(2)\) gauge groups are absolutely
confining, the only difference being that the \(SU(2)\) gauge
force has a scalar field in the elementary representation, and a
choice of gauge where this field is aligned in a fixed direction
is a good point to do perturbation expansion.

It is in \emph{other} Higgs theories where the difference between
the `Higgs mode' and the `confinement mode' is more profound. If
the Higgs were in the adjoint representation, such as in the old
Georgi-Glashow model\cite{GeorgiGlashow} where \(SU(2)\) (without
\(U(1)\)) is spontaneously broken into a \(U(1)\) subgroup by a
Higgs triplet field, then it is not possible to rewrite the
electron or the neutrino, which are in the elementary
representation, as bound states of fermions and scalars.
Nevertheless, electrons and neutrinos are physical particles in
this theory; they are `exotic hadrons', and it is more difficult
to regard them as gauge-invariant objects.
\subsection{Aggregation modes}
Thus, the real question in QCD was: why can quarks not emerge as
physical particles in the same manner as electrons and neutrons do
in the Georgi-Glashow model? The answer to this question is now
known: gauge theories such as QCD and the Georgi-Glashow model
\emph{condense in different aggregation modes}; a system can be
forced to make a transition from one state into another, but such
a transition would necessarily be associated with a \emph{phase
transition}. It is either the electric charges, or the magnetic
charges that can undergo Bose condensation as described in the
above chapters, but never both.

But, to what extent do we \emph{need} the existence of electric or
magnetic charges to realize either one aggregation state or the
other? Could it be that the condensation of the magnetic charges
in QCD is to be seen as a \emph{consequence} rather than the cause
of the confinement mechanism?

The close relation between confinement mechanisms and the
condensation of charges appears to be indisputable. For instance,
it was derived that confinement may occur in an \emph{Abelian}
gauge theory on the lattice. Indeed, this theory also possesses
magnetic monopoles, that appear to condense. In our alternative
treatment of the Standard model, Subsection \ref{Intermezzo.sub},
and notably in the Georgi-Glashow model\cite{GeorgiGlashow}, the
Higgs field is taken to have a large vacuum value, meaning that
these particles Bose condense. Indeed, also, our treatment of
confinement in Section \ref{Abelproj.sec} shows that the
topological argument works for all \(N_c\). However, for large
\(N_c\), the relevant coupling parameter is \(\tl g^2=g^2N_c\),
which means that the electric charges have the strength \(\tl
g/\sqrt{N_c}\), and magnetic charges are combinations of \(g_i\),
with strength \(g_m=2\pi\sqrt{N_c}/ \tl g\). Since the
interactions among these monopoles clearly tend to infinity at
large \(N_c\), treating them using perturbation expansions in
terms of fields becomes questionable.
\subsection{Dynamics}
Therefore, one may argue that, yes, magnetic monopoles do condense
in the confinement mode, even at large values of \(N_c\), but, no,
the actual \emph{mechanism} of confinement could depend on
additional dynamical forces. One expects the hadronic mass scale
at large \(N_c\) to be controlled by its \(\L\) parameter (the
integration constant in the solutions to the Gell-Mann-Low
equation \eqn{betaQED}, \eqn{SU3beta} for the coupling strength),
and this depends on \(\tl g\), not directly on \(g\).

Note that the same arguments could be brought forward concerning
the contributions of instantons . Their action, too, depends on
\(g\) and not \(\tl g\), so that one might expect that they are
exponentially suppressed at high \(N_c\). This is actually known
not to be the case(See for instance Th. Schaefer's contribution at
this Meeting). We do have a \emph{running} coupling strength \(\tl
g\), so that instantons with large sizes are not exponentially
suppressed. Similarly then, one might attribute confinement at
large \(N_c\) to \emph{large} magnetic monopoles.

Large magnetic monopoles would require a fundamentally non-local
effective field theory. The question then remains whether it is
possible to re-establish locality (to some extent) in an effective
local field theory for confinement. A model for that is outlined
in the next section.

\newsec{A classically confining theory} Absolutely confining
forces can indeed be described totally classically. We now
describe such a classical model, also described in
Ref.\cite{GtHQuarks}. It will not be renormalizable, and this
means that, eventually, one wants to attribute the
non-renormalizable terms in the action to quantum effects, so that
at small distances, renormalizability is restored.

Our model contains an Abelian Maxwell field \(A_\m\), and a
neutral, scalar field \(\vv\) that affects the dielectric constant
of the vacuum (in a Lorentz-invariant way): \be \LL(A,\vv)=-\quart
Z(\vv)F_{\m\n}F_{\m\n}-V(\vv)+J_\m(x) A_\m\ , \eel{ConfL} where
the functions \(Z(\vv)\) and \(V(\vv)\) are to be specified later,
and \(J_\m(x)\) is some external source, typically describing
charged `quarks'. We only need its fourth component, the charge
density \(\r(x)\). The scalar field has no kinetic term,
\(-\half(\pa_\m\vv)^2\). We could easily have added that, but it
does not affect the result in any essential manner, and the
calculations are easier when it is (temporarily) ignored.

\begin{figure}[h]
\begin{quotation}
 \epsfxsize=135 mm\epsfbox{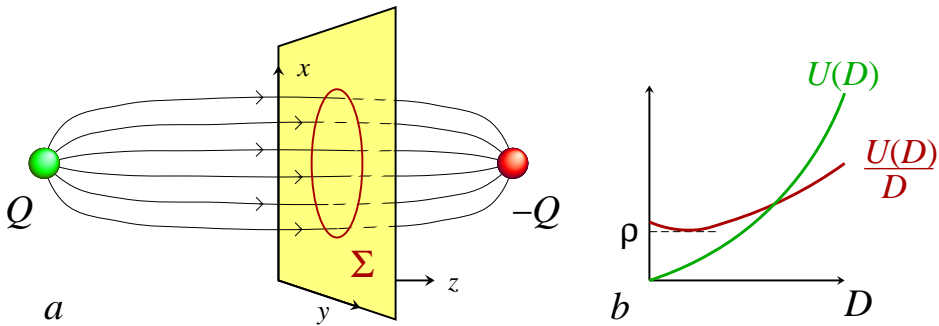}
  \caption{\small{a) vortex spreading out over a surface \(\SS\). b) Graphic
  calculation of the string constant \(\r\).}}
  \label{fluxenergy.fig}\end{quotation}
\end{figure}

\def\bx{\mathbf{x}}
To describe stationary solutions, we use the induction field
\(\vec D(\bx)\): \be \pa_i D_i=\r(\bx)\ ;\qquad D_i=Z(\vv)E_i\
;\qquad E_i=-\pa_i A_0\ . \eel{inductionfield} The Hamilton
density is \be\HH=\half{\vec D^2\over Z(\vv)}+V(\vv)\ ,
\eel{hamilton} and, given the strength of the induction field
\(D\), the energy density \(U(D)\) is obtained by minimizing
\(\HH\) while varying \(\vv\): \be
U(D)=\min_\vv\Bigg(\half{D^2\over Z(\vv)}+V(\vv)\Bigg)\ ;\qquad
D=|\vec D|\ . \eel {Umin} Now consider a field \(\vec D\)
stretching in the \(z\)-direction. Take for simplicity the case
that \(D\) is (more or less) constant over a surface \(\SS\)
stretching in the \(xy\) direction, see Fig.
\ref{fluxenergy.fig}$a$. Because of \eqn{inductionfield}, \(\vec
D\) represents a total charge \(Q=D\SS\). So, suppose that the
surface area \(\SS\) is allowed to expand to any arbitrary size.
Then the minimal energy per unit of length is \be
\r^\mathrm{string}=\min_\SS\Big(\SS U\big({Q/\SS}\big)\Big) =
Q\,\min_D\Big({U(D)\over D}\Big)\ . \eel{rhostring}

So, \emph{if} \(U(D)/D\) has a minimum \(\r\), preferably at some
finite value of \(D\) (see Fig.~\ref{fluxenergy.fig}$b$), then we
see that a vortex emerges, with string tension \(\r\), spreading
out more or less evenly over the surface \(\SS\), while the \(D\)
field tends to zero outside this surface. This condition is met if
\(U(D)\) is linear in \(D\) for small \(D\) (unlike the Maxwell
case, where \(U(D)=\half D^2\)). In Eq.~\eqn{Umin}, this is
realized if \be Z(\vv) \simeq C\cdot V(\vv)\ ,\qquad C=2/\r^2\ ,
\eel{VZrel} near the minimum of \(V\).

\begin{figure}[h]
\begin{quotation}
 \epsfxsize=135 mm\epsfbox{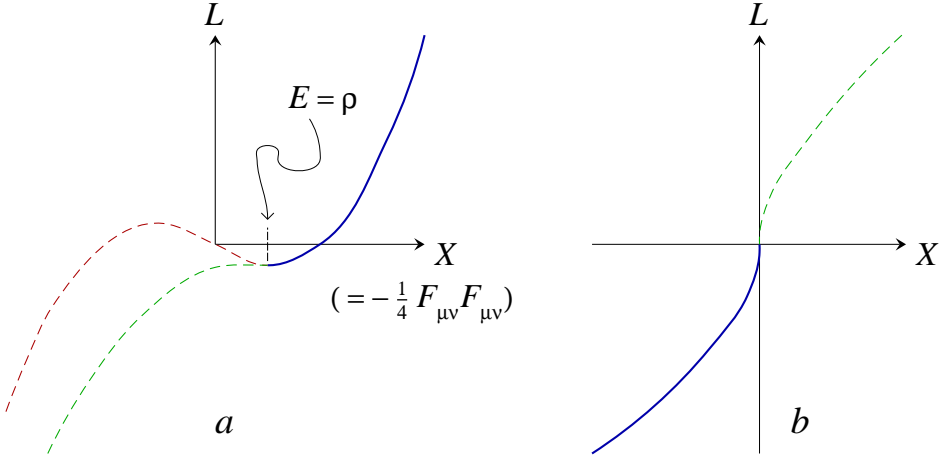}
  \caption{\small{$a)$ The Lagrangian \(L\) as a function of the quantity
  \(X=-\quart F_{\m\n}F_{\m\n}\,\), in the case of electric
  confinement. The (blue) solid line shows behavior necessary for
  confinement. The dotted lines show different allowed
  continuations. $b)$ The dual of $a)$, the magnetic confinement
  case.}}
  \label{Lcurves.fig}\end{quotation}\end{figure}

It is easy, also, to guess the effect of a possible kinetic term
for \(\vv(x)\), which we had ignored. It will only contribute at
the surface of this (finite size) vortex, so that the \(D\) field
will not show \(\th\) jumps at the edges of \(\SS\), but grow more
smoothly from zero outside, to the fixed value \(D\) inside the
vortex.

If we leave out the kinetic term for \(\vv(x)\) altogether, then
we may just as well eliminate \(\vv\) from the Lagrangian
\eqn{ConfL} at the very beginning. Assuming Eq.~\eqn{VZrel} for
small \(V\), and \(Z\simeq 1\) for large \(V\) (so that the
small-distance structure of the theory tends to the renormalizable
situation), we find that the effective Lagrangian as a function of
the entry \be X=-\quart F_{\m\n}F_{\m\n}=\half \vec E^2\ ,
\eel{entry} is obtained from the equations \be X={\dd V\over\dd
Z}\ ,\qquad \LL(X)=ZX-V\ . \eel{LLx} The curve is depicted in Fig.
\ref{Lcurves.fig}$a$.

 The magnetic confinement case is obtained by replacing \(D\) with
 \(B\), and \(Z\) with \(1/Z\) in Eqs.~\eqn{hamilton} and
 \eqn{Umin}. In that case, only the negative values of \(X\)
 count, and the required behavior of the Lagrangian is depicted in
 Fig.~\ref{Lcurves.fig}$b$.

 Investigating various functions \(Z(V)\) is an instructive
 exercise. Further explanations can be found in
 Ref.\cite{GtHconf}.

\end{document}